\documentclass[%
nofootinbib,
 amsmath,amssymb,
 aps,
]{revtex4-2}

\usepackage{graphicx}   
\usepackage{verbatim}   
\usepackage{color}      
\usepackage{subfigure}  
\usepackage{hyperref}   
\usepackage{soul}
\usepackage{hyperref}
\usepackage{tabularx}

\usepackage{float}
\usepackage{amssymb}
\usepackage{mathtools}
\usepackage{setspace}
\usepackage{amsmath}
\usepackage{bm}
\usepackage{outlines}
\usepackage{xcolor,cancel}

\usepackage[utf8]{inputenc}
\usepackage[T1]{fontenc}
\usepackage{mathptmx}
\usepackage{lmodern} 

\usepackage{dcolumn}
\usepackage{float}

\usepackage{tcolorbox}
\usepackage{graphicx}
\usepackage{multirow}
\newcommand{\nn}{\textcolor{black}}
\begin{document}

\preprint{APS/123-QED}

\title{Compression-controlled dynamic buckling in thin soft sheets}

\author{Harsh Jain} \email{harshjainldh@gmail.com}
\affiliation{%
Department of Condensed Matter Physics and Materials Science,\\ Tata Institute of Fundamental Research, Mumbai 400005, India}

\author{ Shankar Ghosh}
\affiliation{%
Department of Condensed Matter Physics and Materials Science,\\ Tata Institute of Fundamental Research, Mumbai 400005, India}

\author{Kirti Chandra Sahu}
\affiliation{%
Department of Chemical Engineering, Indian Institute of Technology Hyderabad, Sangareddy, Telangana, 502 285, India
}%


\begin{abstract}
We investigate experimentally the dynamic phase transition from compressed to buckled phases for thin sheets of rubber. We find that the rubber strips enter a highly compressed, meta-stable state when compression speed is high. During the compressed phase, higher modes grow, followed by mode-coarsening. Mode growth is accompanied by an expansion of length while the system is still being compressed.  We measure the forces and length of the sheet to confirm this, and we develop a mechanism for how modes grow and coarsen during dynamical buckling.   The influence of crucial control parameters in the experiments, such as the material cross-section and compression speed, on the buckling dynamics, are explained theoretically. 
\end{abstract}

\maketitle


\section{Introduction} \label{sec:intro}

Buckling is an abrupt change in shape (deformation) of a structural component under load/compressive stress. In a buckled condition, which occurs when compressive stress surpasses a critical value \cite{landau1987elasticity}, the stress and strain are no longer co-linear. The value of this critical stress is vanishingly small for slender structures. Therefore, thin sheets or rods buckle easily when a compressive force is applied at their ends while thicker materials prefer getting compressed. This is in line with our everyday experience: a rubber eraser can be compressed while a rubber band buckles (Figure \ref{fig:exptsetup}(a)). Buckling is a common occurrence in a wide range of systems, from engineering to biology, and over a wide range of length scales \cite{wu2020harnessing,hu2017axial,jiang2008mechanics}. 

In addition to structural engineering, the buckling phenomenon has applications in a vast variety of modern technologies. The dissipated energy from buckling can be used in absorbers, dampers, stabilisers, and isolators in energy-related applications, while the energy generated during buckling events can be used to design energy harvesters and sensors for micro-electromechanical systems (MEMS). Buckling and post-buckling responses have also been identified as useful design aspects in smart gadgets and mechanical systems. A comprehensive review of several applications in modern technology can be found in \citet{hu2015buckling}. 

A system remains in an equilibrium state at all times when the control parameter (loading) is altered quasi statically; this is known as ``static buckling''. Here,  buckling in the fundamental mode is supposed to be an energetically favourable mode of deformation. However, in films and membranes, the presence of a substrate or medium can occasionally promote the growth of higher-order modes \cite{cerda2003geometry,broedersz2014modeling}. When the control parameter is altered quickly enough that the system does not have time to equilibrate, and the buckling process is referred to as ``dynamic buckling'' \cite{lindberg2003little,vermorel2009impacts,vandenberghe2013geometry}. 

In this study, we experimentally show that under suitable conditions, a thin free-standing rubber sheet can be dynamically driven deep into a metastable compressed state. We establish this by concomitantly measuring the compressive forces and the instantaneous length of the strip. It is found that a  transition from this compressed to the buckled state is accompanied by a sudden increase in the length (negative compressibility). In order to get the physical insights, we solve a fourth-order dynamic buckling equation and use a three-spring model to explain the scaling behavior observed in the experiments. To our knowledge, no earlier studies \cite{chopin2017dynamic,oratis2019shooting,kodio2020dynamic}  of dynamic buckling of thin filaments in the presence of a surrounding media or during free retraction (i.e., stretch and let go) of rubber have described this behaviour.

\section{Experimental Method} \label{sec:expt}

\begin{figure}[t]
	\centering
	\includegraphics[width=.9\linewidth]{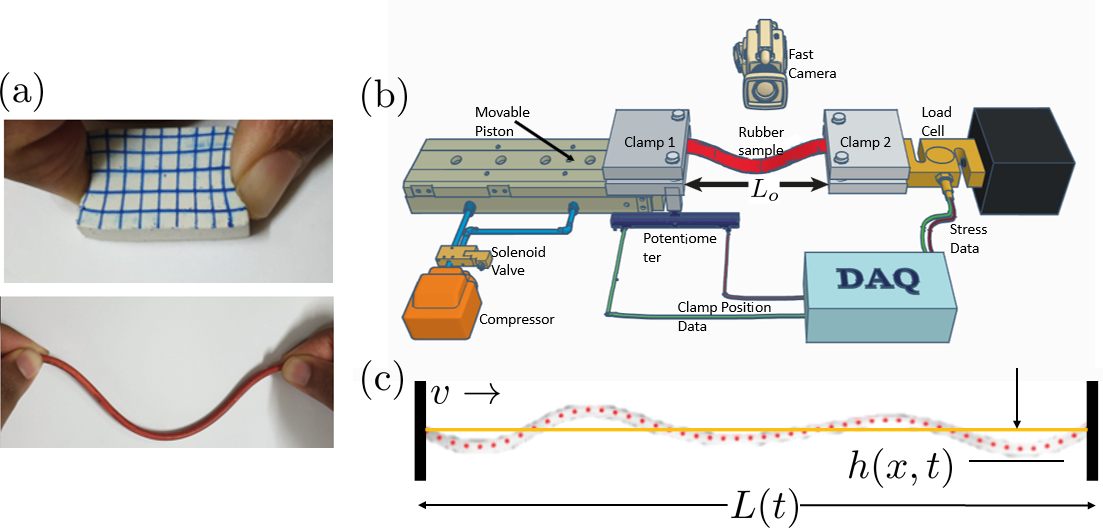}
	\caption{{Demonstration of the key idea and experimental realisation of the buckling.} (a) Demonstrates that while an eraser gets compresssed, a rubber band buckles easily. (b) A schematic representation of the experimental setup. (c) Image of a thread held between two clamps with the left clamp moving. The red dots show the detected points used to calculate the contour length  $\ell (t)$ of the strip. Here, $h(x,t)$ and $L(t)$ denote the height field of the strip and the distance between the clamps. The red points along the strip's length are representations of the central maxima of the strip's image along its vertical section. Connecting these points yields the contour along the length of the strip.}
	\label{fig:exptsetup} 
\end{figure}

The schematic diagram of our experimental setup, which includes a rubber sample, two clamps, a load cell, a movable piston arrangement, a solenoid valve, compressor, a high-speed imaging system, and a data-gathering system, is shown in Figure \ref{fig:exptsetup}(b). We move the left clamp inwards while holding a stretched thin strip of rubber between two clamps. The right clamp is fastened in place and connected to a $50$N S-type load cell, which measures the instantaneous forces applied on the rubber strip during unloading. The left clamp is connected to a Festo DGSL pneumatic linear slide with a solenoid valve switch that controls the direction of motion. Both clamps are mounted on mechanically isolated tables to eliminate vibrations or jerks caused by the pneumatic slide's movement from interfering with force measurements. The slider's speed can be varied from 1 m/s to 0.01 m/s by adding external springs in parallel that resist its motion.  Except for the first few milliseconds when it starts moving and the last few milliseconds when it reaches its final location, the pneumatic slide's pace is uniform. An analogue linear potentiometer with a sensitivity of 0.27 V/mm is used to measure the slide's location. An integrated circuit chip was used to amplify the electronic signals from the load cell (INA110KP). At time intervals of $20\mu$s, all electronic signals were continuously captured with a National Instruments Data Acquisition (NI-DAQ) system. We observed the side view of the strip's motion with a Phantom Miro M310 camera at a consistent interval of $\sim 66\mu$s. The resolution of these photos is 100 $\mu$m/pixel. The electronic and image data are synchronised with a precision of more than 0.1 ms in time to observe the forces corresponding to the strip changing configuration. The timestamps of the electronic signal from the potentiometer and the image of the slider as detected by the camera were aligned to achieve this. 

To obtain its instantaneous length, the contour length $\ell$ along the centre of the strip from the left to right clamp is measured.  Here, $\ell (t) = \int  dx \sqrt{1+(\partial{h(x)}/\partial{x})^2}$, with  $h(x,t)$  being the  out-of-plane height fluctuation and  $x$ indicating the direction along the straight strip.  Figure \ref{fig:exptsetup}(c) shows a typical lateral image recorded (from the side). Connecting the central maxima of the strip's image along the vertical section yields the contour along the length of the strip. Fitting the appropriate vertical profile of the image to a Gaussian function yielded this central maximum at a given point.  A subset of  these  detected  points  are shown as red dots  along  the length of the strip in Figure \ref{fig:exptsetup}(c). The MATLAB\textsuperscript{\textregistered}  code for the picture analysis can be found in the GitHub repository \cite{Jain2021}. This approach presupposes two-dimensional dynamics, which we found to be a valid assumption in our experiments. However, this assumption is temporarily violated when the slider comes to a standstill and twisting modes appear in the strip.   Rubber strips with an initial length of $ \ell_o $, a width of $ w_o $, and a thickness of $ \zeta_o$ are cut from seamless valve bladders manufactured of acrylonitrile-butadiene rubber (thickness 1 mm) and natural rubber latex bands (thickness 0.2 mm) widely accessible as exercise bands. Both are filled-type rubbers that demonstrate stress relaxation and residual strain relaxation on a time scale of $\sim 1$ s \cite{dargazany2013constitutive,diani2009review}, which is significantly shorter than the timescale of our studies ($\sim 1$ ms). Thus, the gravitational impact can be ignored. The elasticity of these rubber samples was measured to be about 3 MPa.

\section{Results and Discussion} \label{sec:dis}

\begin{figure}[t]
	\centering
	\includegraphics[width=0.95\linewidth]{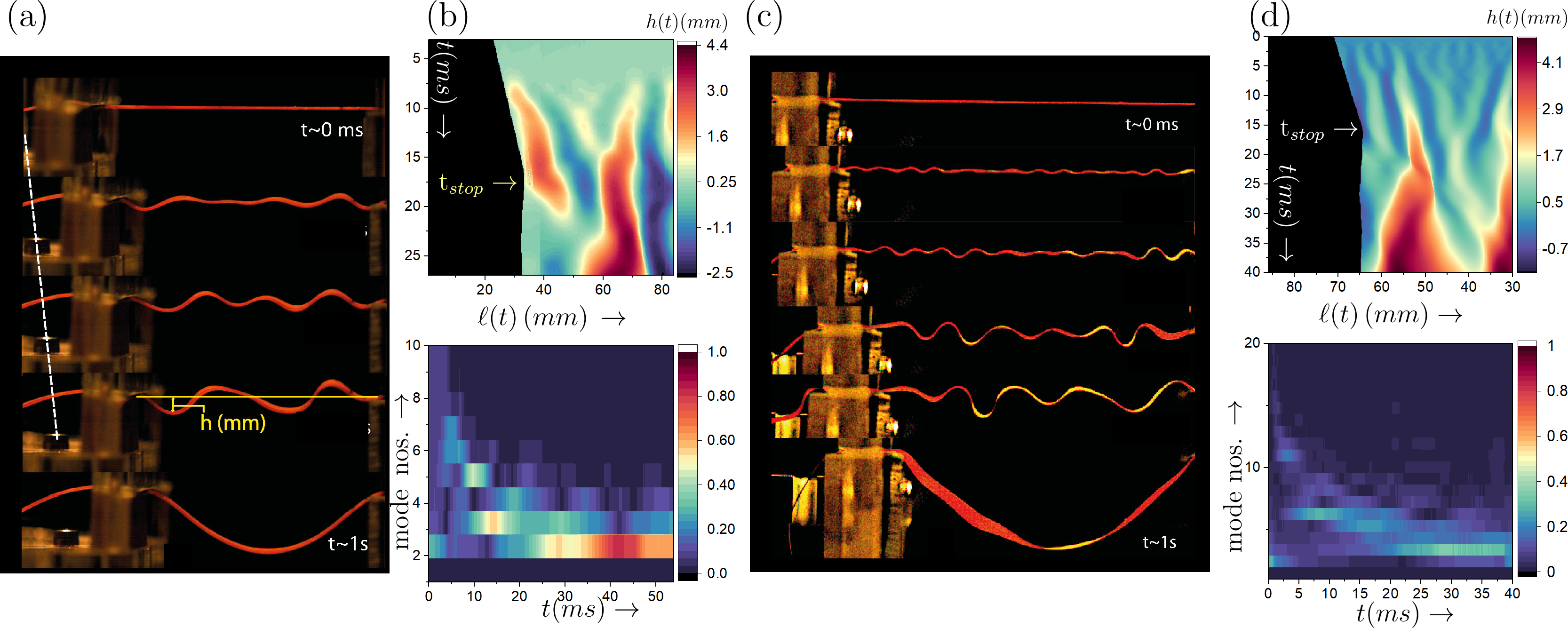}
	\caption{{Spatio-temporal evolution of the mode structure during buckling.} (a)-(b) show data for a strip of acrylonitrile rubber (dimension: initial length $\ell_o=$ 86 mm, width $w_o=$ 5 mm, and thickness  $\zeta_o=$1 mm) compressed at speed $v =1$ m/s. The panel (a) presents snapshots that capture the process of buckling. The top panel of (b) shows the spatiotemporal evolution of the height map $h(x,t)$ and  the bottom  panel shows the time evolution of the fractional amplitude $h_k/\sum_{n}h_{n}$ of modes  obtained from the Fourier decomposition of $h(x,t)$.  Here, $h$ is taken to be positive in the direction of gravity. (c)-(d) depict the corresponding results for the natural rubber (dimension: $\ell_o= 80$ mm,  $w_o= 5$ mm and $\zeta_o=0.2$ mm). }\label{waterfallPlot}
\end{figure}

Figures \ref{waterfallPlot}(a-b) and (c-d), respectively, show the buckling dynamics of acrylonitrile and natural rubber strips. Representative images of the dynamic buckling in acrylonitrile (1 mm thickness) and natural rubber (0.2 mm thickness) are shown in Figures \ref{waterfallPlot}(a) and \ref{waterfallPlot}(c). When buckling begins, higher-order wrinkles form and the system advances to the fundamental mode over time. As demonstrated in Figures \ref{waterfallPlot}(b) and (d), higher-order modes cascade down to the lowest-order mode as time progresses.

The out-of-plane height fluctuation $h(x,t)$ over the length of the strip ($x$) is shown in the contour plots in the top panel of Figures \ref{waterfallPlot}(b) and (d). At a moment labelled as $t_{stop}$  in Figure \ref{waterfallPlot}(b), the clamp reaches its final position and the system is dramatically disrupted. The time evolution of the amplitude of the various modes is depicted in the bottom panel of Figures \ref{waterfallPlot}(b) and (d). The mode structure coarsens over time, eventually reaching $n=1$ (fundamental mode) as shown in Figures \ref{waterfallPlot}(b,d). \citet{kodio2020dynamic} developed an analytical expression for the stability of all the modes of order $n$ when the dynamic buckling occurs under a constant compressive force. They demonstrated that the maximum allowed $n$ scales with the square root of the compressive force. 
As compressive force increases in our system, the maximum $n$ observed in our system also increases (see Fig. 2 (b)).
However, for a time varying force, it's possible that the mode selection criteria will change from that given in  \citet{kodio2020dynamic}.

\begin{figure}[t]
	\centering
	\includegraphics[width=0.95\linewidth]{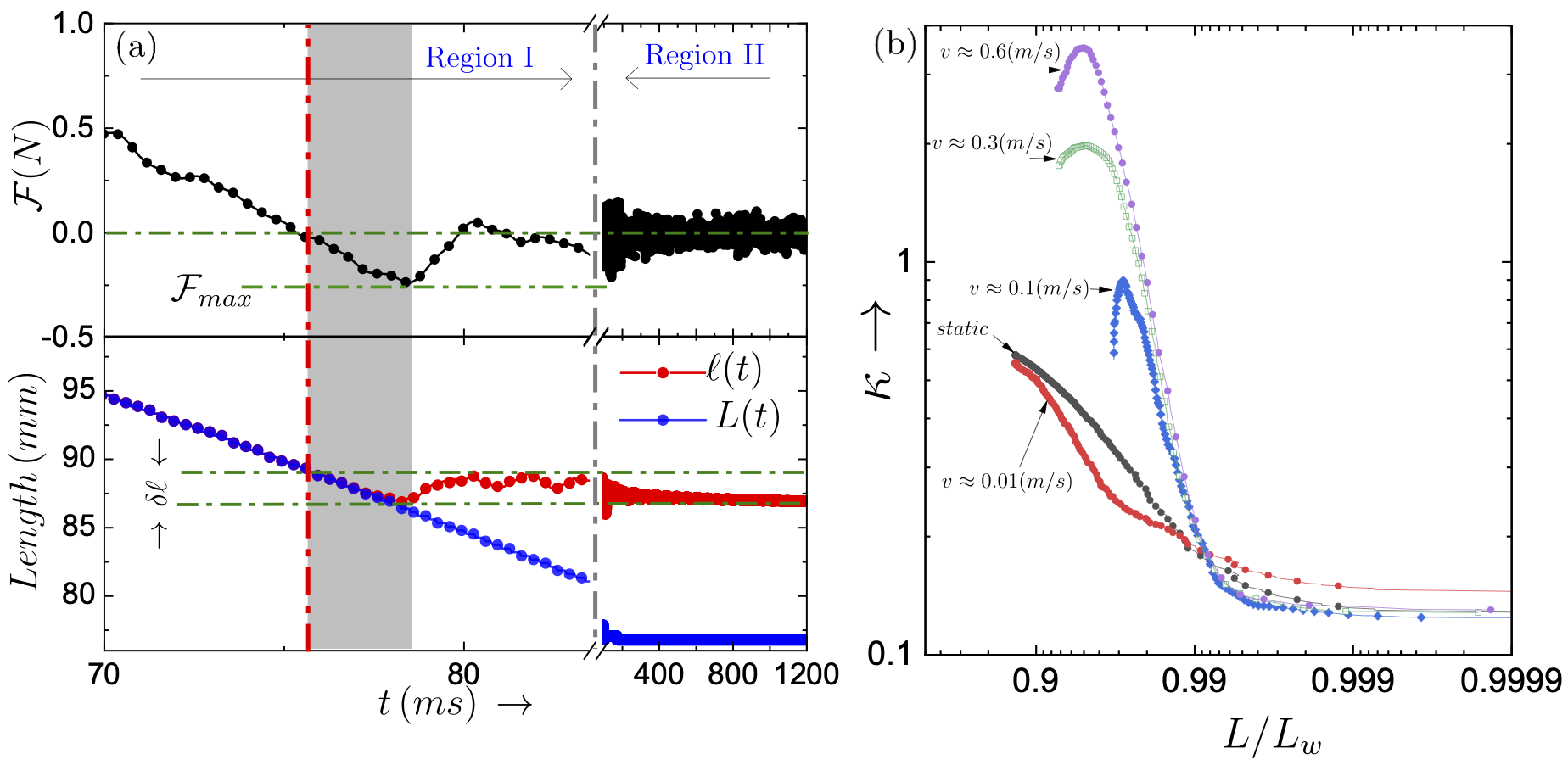}
	\caption{{Structural response during buckling and evidence of negative compressibility.} (a) The temporal variation of the measured force $\mathcal{F}$ (top panel), the contour length $\ell(t)$ (\nn{upper curve in the bottom panel}) and distance between clamps $L$ (\nn{lower curve in the bottom panel}). Here, Region I shows the subtle features near the instance of buckling, and Region II shows the long-time evolution. (b) Variation of the total absolute curvature of the strip $\kappa$ as a function of time, for various values of the compression speed \nn{(marked against each curve)}. In these experiments, $\ell_o = 86$ mm and $L_o=78$ mm.}
	\label{fig4:StressLength}
\end{figure}

Figure \ref{fig4:StressLength}(a) depicts the temporal change of the force $\mathcal{F}$, instantaneous distance between clamps $L(t)$, and instantaneous contour length $\ell(t)$ of the rubber strip observed during the experiment.  Region I displays the variation surrounding the buckling site in Figure \ref{fig4:StressLength}(a), while the long-term evolution of the strip's shape is depicted in Region II. The clamping distance is then raised to  a maximum distance of $\approx 108$ mm. This causes the strip to become taut. The amount of stretching that occurs at the start has no significant impact on our experimental results. We externally control the retraction pace using the springs attached to the clamp to ensure that it travels at a constant speed $v$, unlike in free retraction experiments \cite{oratis2019shooting} where the initial stretching determines the speed of retraction.

In the grey region shown in Figure \ref{fig4:StressLength}(a), the force becomes compressive. Here, both the higher-order modes and the compressive force grow in amplitude. The length of the strip thus begins to extend as these modes proliferate, while the compressive force $\mathcal{F}$ continues to rise, indicating that the material transiently shows negative compressibility. The expansion of these modes is limited by the strip's finite elasticity. As the strip reaches the equilibrium length $\ell_{eq} \approx 88$ mm, the modes coarsen and the compressive force $\mathcal{F}$ in the system goes to zero. Similar force relaxation has also been recently observed by \citet{villermaux2021mode}. 
This equilibrium length decays over a period of $1$s in Region II, returning to the length $\ell_o$ of the unstretched strip of $\sim 86$ mm via residual the strain relaxation process \cite{roland2006mechanical}. This period is far longer than the timescale of the buckling dynamics, which is of the order of a few milliseconds.  It's worth noting that the oscillations visible at the start of Region II are the consequence of the system's twisting modes being generated by the slider coming to a sudden stop.

The height field $h(x,t)$ evolves in a spatio-temporal manner as the strip buckles. To study its temporal evolution, we apply a trigonometric series expansion of the height field 
\begin{equation}
h(x,t) = \sum_k h_k(x,t) \sin(kx),
\label{Eqn:Fourier}
\end{equation}
where $k = n\pi/L$ is the wavenumber $(n=1,2,3 \cdots)$ and $L$ specifies the distance between ends of the strip. Here, $h_ {k}(x,t)$ will have the form $\sim e^{\alpha(x,t)}$ due to the moving boundary.
 
Figure \ref{fig4:StressLength}(b) shows the total absolute curvature, $\kappa=\int \left|\frac{\partial_{xx}h(x,t)}{(1+\partial_{x}h(x,t)^2)^{1/2}}\right| dx$ at the onset of buckling for the same strip compressed at different speeds, with $L/L_w$ on the $x$ axis, wherein, $L_w$ is the distance between clamps concomitant with the onset of buckling. Qualitatively, the curves corresponding to the dynamic $(n>1)$ and static $(n=1)$ buckling are very different. The early part of the increase of $\kappa$ for driving at different speeds follows the same pattern in the dynamic buckling example. The value of $\kappa$ increases as the compression speed increases, before decaying back to the curvature associated with the static buckling condition.

\begin{figure}[!t]
 	\centering
 	\includegraphics[width=.95\linewidth]{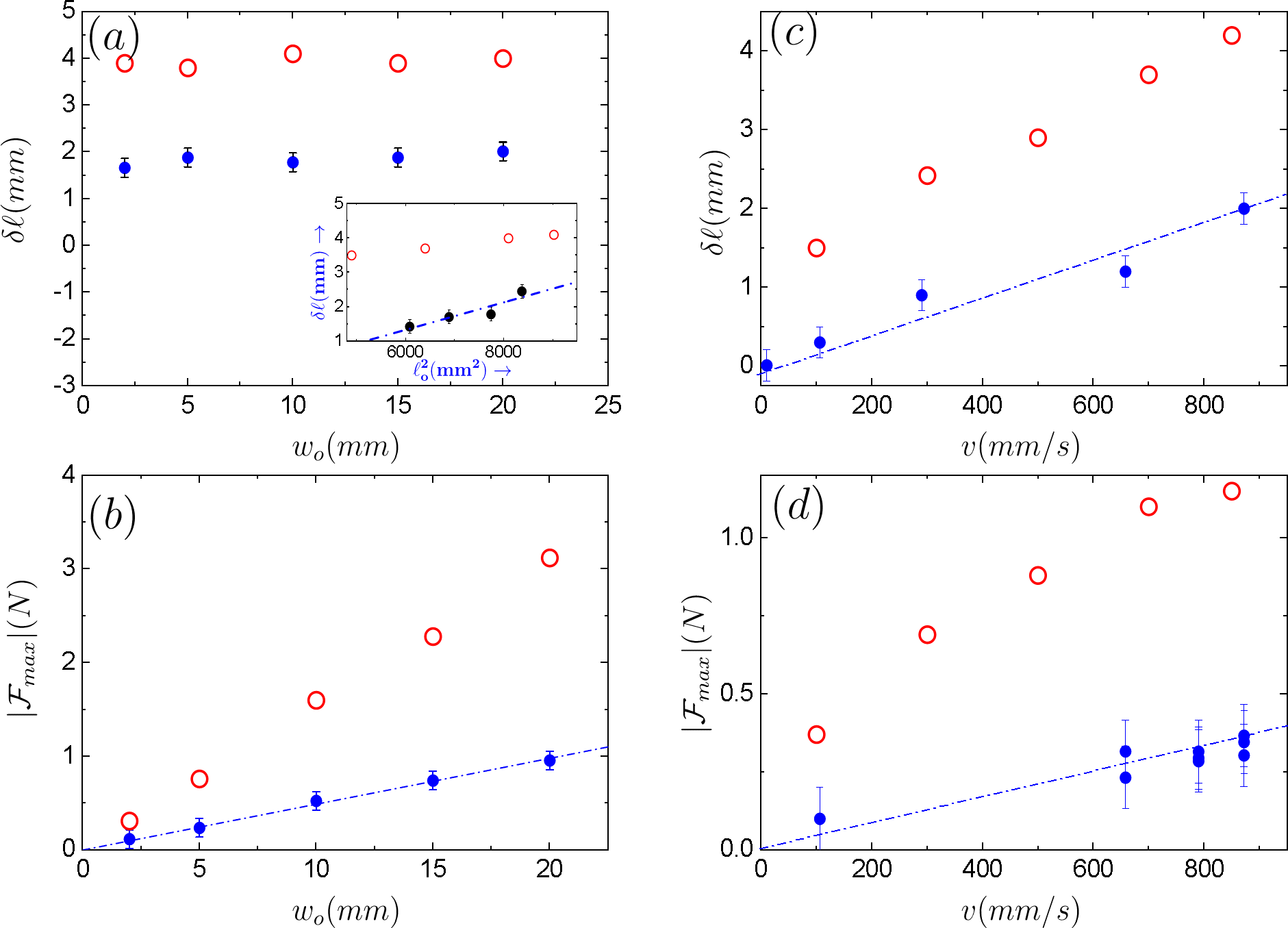}
 	\caption{{Observed scaling laws.} (a) and (b) show the variation of the jump length $(\delta \ell)$ and the maximum force sustained by the strip $(\mathcal{F}_{max})$, respectively with the width of the sample $(w_o)$. The panels (c) and (d) show the variation of these parameters with the compression speed $v$. Here, $\ell_o \sim 8$ cm, thickness $\zeta_o = 1$ mm for (a)-(c) and $\ell_o \sim 15$ cm,$\zeta_o=1$ mm  for (d). In panels (a) and (b), $v \approx 0.8$ m/s and in panels (c) and (d) $w_o=10$ mm. The corresponding numerical simulation results obtained by solving Eq.(\ref{Euler-Bernoulli2}) are presented as red open circles with $ I= {w_o \zeta_o^3/12}$  and $A= w_o \zeta_o$. The  simulation parameters  are  same as experiments, with  $\rho= 1.1$ kg/m$^{3}$ and  $G= 3$ MPa. \nn {The  dash-doted lines  in the figures represent the various scaling laws 
  which are  	obtained from  Eqs. (\ref{DeltaVar}) and (\ref{FmaxVar}). These scaling laws are (i) $\delta \ell$ is independent of $ w_o $ (ii) $\delta \ell \propto \left| v\right|  \ell_o^2$, and (iii)  $\left|\mathcal{F}_{max} \right| \propto  w_o \left|v\right|$.} } 
 	\label{parameterVariation}
\end{figure}

The variation of the jump in the length $\delta \ell$ and the magnitude of $|\mathcal{F}_ {max}|$ as a function of the sample width and compression speed $v$ is depicted in Figure \ref{parameterVariation}. The key observations from this result are that $\delta \ell$ is unaffected by strip width $w_o$ and that it varies linearly with $v$ and $\ell^2$. Furthermore, with both width $w_o$ and speed $v$, the absolute value of $\mathcal{F}_{max}$ rises linearly.

\begin{figure}[t]
 	\centering
 	\includegraphics[width=1.0\linewidth]{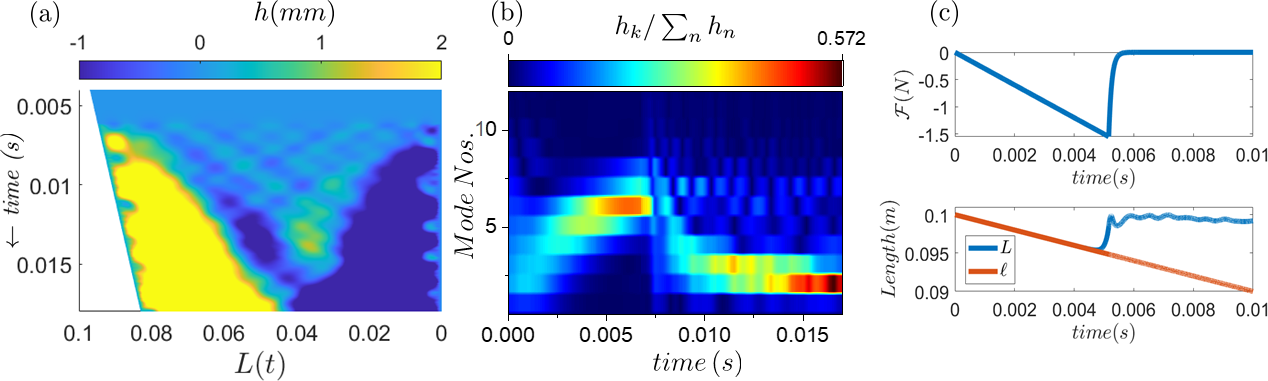}
 	\caption{{Numerical results obtained using dynamical buckling equation.} (a) Spatio-temporal variation of height, $h$ obtained numerically by solving Eq. (\ref{Euler-Bernoulli Equation}).  (b) The temporal evolution of mode fraction $h_k/\sum_n h_n$. The top and the bottom panels of (c)   depict the temporal variations of $\mathcal{F}(t)$, and the contour length $\ell(t)$ of the strip, respectively. Here, the parameters are the initial length, $\ell_0 = 86$ mm, thickness $\zeta_o=1$ mm, $w_o = 10$ mm, $\rho= 1.1$ kg/m$^3$ and $G = 3$ MPa.}
 	 \label{Fig:NumericalODE}
\end{figure}

In order to get the physical insights of the dynamics observed, we use the following dynamic buckling one-dimensional (1D) equation for a material with density $ \rho $, cross-section $ \mathcal{A} $ that is subjected to a time varying compression force $\mathcal{F}(t)$:
\begin{equation}
\rho \mathcal{A} {\partial^2 h \over \partial t^2}=-G I {\partial^4 h \over \partial x^4}- \mathcal{F}(t) {\partial^2 h \over \partial x^2},
\label{Euler-Bernoulli Equation}
\end{equation}
where $x$ varies between $0$ and $L(t) = L_o-vt$. Note that, in general, the compression force will also depend on $x$, but we have neglected this variation. In our experiments, the speed of the compression $(v)$ is much smaller than the speed of sound $(c)$; thus this assumption is justified. Using the Galilean transformation by introducing $\tilde{x} = x/L(t)$ that varies between 0 and 1, we can re-write Eq. (\ref{Euler-Bernoulli Equation}) as
\begin{equation}
\rho \mathcal{A} \frac{\partial^2 h}{\partial t^2}=-\frac{G I}{L(t)^4} \frac{\partial^4 h} {\partial \tilde{x}^4}- \frac{\mathcal{F}(t)}{L(t)^2} \frac{\partial^2 h}{ \partial \tilde{x}^2}. \label{Euler-Bernoulli2}
\end{equation}
This fourth-order differential equation is linear in $h$ and the solution will be the sum of many buckling modes that can be Fourier decomposed with boundary conditions $h(0,t)=h(L,t) = \frac{\partial^2 h}{ \partial x^2} \vert _{\left(x=0,L\right)}=0$. The time-dependent force $\mathcal{F}(t)$ leads to the coupling of modes, thus in such cases, Eq. (\ref{Euler-Bernoulli2}) can be solved numerically using {\it ``ode23s''} solver in MATLAB\textsuperscript{\textregistered} \cite{Jain2021}. Figure \ref{Fig:NumericalODE} depicts the numerical results for the initial  length, $\ell_0 = 0.1$ m, thickness $\zeta_o=1$ mm, width $w_o = 10$ mm, $G = 3$ MPa obtained by solving Eq. (\ref{Euler-Bernoulli Equation}) under dynamic loading using the method of lines \cite{wouwer2014simulation}. The set of parameters considered in Figure \ref{Fig:NumericalODE} are the same as the experimental settings. Figure \ref{Fig:NumericalODE}(a)  and (b) shows the spatio-temporal variation of height, $h$  and the growth of the modes obtained numerically by solving Eq. (\ref{Euler-Bernoulli Equation}), respectively.
For the force profile shown in Figure \ref{Fig:NumericalODE}(c) (top panel), we find the modes to grow and coarsen as observed in our experiment (Figure \ref{fig4:StressLength}(a)). While modes grow as long as the compressive force is present, as the strip reaches back to its uncompressed length, the force quickly decays towards zero, and coarsening follows. From this point, we observe oscillations in our experimental force measurement. The numerical solutions also show similar contraction and expansion in contour length $\ell(t)$ of the strip as shown in Figure \ref{Fig:NumericalODE}(c) (bottom panel). Though the numerical solutions agree qualitatively with the  observations shown in Figure \ref{fig4:StressLength}, it overestimates the parameters. The detailed comparison is given in  Figure \ref{parameterVariation}. This disparity can be explained by the fact that the numerical computations were done for a 1D system, whereas the experimental observations are in higher dimension.

\begin{figure}[t]
	\includegraphics[width=0.95\linewidth]{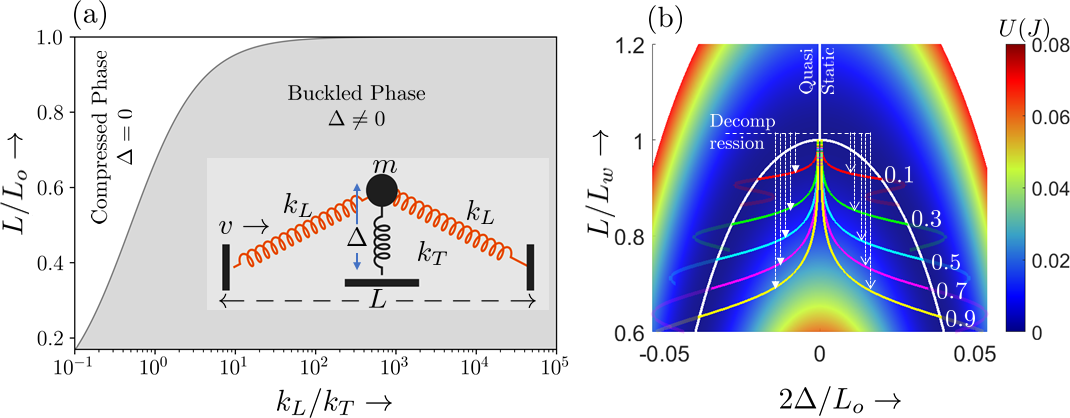}
	\caption{{Results from the three-spring model.} (a) The boundary separating the buckled and compressed phases in the $ (k_L/k_T, L/L_o) $ parameter space. The inset shows the three-spring model. (b) The contours of the potential $U~(2\Delta/L_o,L/L_w)$. Here, the white line represents the static buckling while colored lines show numerical solution for the dynamic buckling at various speeds. The speeds  in units of m/s are marked beside each line. The parameters are $m = 1$ g, and $k_L,k_T$ values are obtained from Eq. (\ref{geometricSpringConstants}) using elasticity $G = 3$ MPa, $w_o = 5$ mm, $\zeta_o=1$ mm and $\ell_o = L_o= 100$ mm . These values are close to the experimental parameters as shown in Figure \ref{fig4:StressLength}. \nn{The  white dotted arrows in panel (b) indicate the  paths by which the springs decompress.}}
	\label{Fig:Fig1}
\end{figure}

Further, we use a toy three-spring model system as shown in Figure \ref{Fig:Fig1} to understand the phenomenon observed in our experiments. Below first, we discuss the system considered and then make a connection of the three-spring model with our experimental results. Figure \ref{Fig:Fig1}(a) shows the phase diagram for the compressed and buckled states obtained from the three-spring model representing a dynamic buckling scenario that captures the essence of the buckling transition for a mechanical column \cite{hutchinson1966dynamic,kuzkin2015structural}. It consists of three hinged springs acting on concentrated mass $m$. Two of these are longitudinal springs with spring constant $k_L$ that connect the mass with the clamps. The left clamp is driven inwards with speed $v$ while the right clamp stays fixed. These springs whose equilibrium length is $a_o$ oppose the compression forces that are generated when the clamp is driven inwards. The third spring with spring constant $k_T$ whose equilibrium length is `zero' is so arranged that it opposes the traverse displacement $\Delta$. This spring captures the column's resistance to bending. Here,  $L$ is the instantaneous end-to-end distance between the clamps and $L_o$ is the equilibrium length. The potential energy of this system is $U=k_L(\sqrt{(L/2)^2+\Delta^2}-(L_o/2))^2+k_T\Delta^2/2$. Solving for the minimum potential energy by setting $dU/d\Delta =0$, we find that the system undergoes a transition from compressed phase to buckled phase at the critical end-to-end length $L=L_w$, where $ L_w=L_o k_L/(2k_L+k_T)$. In the parameter space $ (k_L/k_T, L/L_o) $, this transition generates a boundary between buckled (light gray, $ \Delta \ne 0$) region and the compressed (white, $ \Delta = 0$) region in Figure \ref{Fig:Fig1}(a). The contours of the potential $U~(2\Delta/L_o,L/L_w)$ is shown in Figure \ref{Fig:Fig1}(b), wherein $U$ represents the calculated value at each point in the parameter space of $2\Delta/L_o$ and $L/L_w$. The white lines represent Min$\{U(L/L_w,2\Delta/L_o) \}$ during quasi-static compression (static buckling). At $ L=L_w $, a pitchfork bifurcation \cite{pippard1985response} happens. For $\left(L/L_w\ge1\right)$, $\Delta =0$ (compressed state) and  for $\left(L/L_w<1\right)$, $2\Delta =\pm\sqrt{L_w^2-L^2}$ (buckled state) are the positions of stable equilibrium.

The dynamical aspect of this phase transition can be captured by writing down the Lagrange's equation in the generalised coordinate $\Delta$ as 
\begin{equation}
  m {d^2\Delta \over dt^2} = -{dU \over d\Delta} 
                    = \left[2k_L(1-L_o/L)+k_T\right]\Delta.
\label{Lagrange Equation}
\end{equation}
The coloured lines in Figure \ref{Fig:Fig1}(b) show the solution of this equation for different compression speeds ($dL/dt$). It shows that the system overshoots at the critical point $L/L_w = 1$ before undergoing the transition. The extent of this overshooting increases with increase in the compression speed. \nn{It is to be noted here that the oscillations in the color lines correspond to the vibration (ringing motion) of the springs about their equilibrium values. This is similar to the    vibrations generated in the rubber strip  following the phenomenon of sudden expansion (see the red-dot curve in the bottom panel of Fig. \ref{fig4:StressLength}(a)).} The spring constants are related to the geometrical descriptors of the column (length $\ell_o$, width $w_o$ and thickness $\zeta_o$) in the following way:
\begin{equation}
k_L= \frac{G w_o \zeta_o }{\ell_o} \quad \mathrm{and} \quad k_T= \frac{G I}{ \ell_o^3},
\label{geometricSpringConstants}
\end{equation} 
where $I= {w_o \zeta_o^3/12}$ and $G$ are the moment of inertia and elastic constant of the column, respectively. 

Equation (\ref{Lagrange Equation})  describes the motion of a harmonic oscilator with 
a characteristic time scale $ \tau =\sqrt{m/k_{eff}}$, wherein $ k_{eff} = \left[2k_L(1-L_o/L)+k_T\right]$. During  this time, the system keeps moving on any one of the colored dashed lines (Figure \ref{Fig:Fig1}(b)) and makes a transition to the buckled state at $ L<L_w $.
An excess energy $ \delta U \approx \frac{1}{2}k_Lv^2 \tau^2= \frac{1}{2}(k_L/k_{eff})mv^2$ in form of compression is stored in the longitudinal springs. This energy is released during the process of buckling. During this process, the longitudinal springs decompress by a distance $\delta\ell= v\tau \approx v\sqrt{m/k_{eff}} $. Therefore, from the spring model with $c = \sqrt{{G/\rho}}$, we obtain
\begin{equation}
\delta \ell =\frac{v}{c}\sqrt{\frac{\ell_o^4}{2\ell_o^2\left(1-\frac{L_o}{L}\right)+\zeta_o^2}}
\label{DeltaVar},
\end{equation}
 and the maximum compression force to be 
\begin{equation}
     \left|\mathcal{F}_{max}\right|=  \frac{1}{2}k_L \tau \left| v\right| =
      \frac{\left| v\right|G w_o \zeta_o}{2c\ell_o} \sqrt{\frac{\ell_o^4}{2\ell_o^2\left(1-\frac{L_o}{L}\right)+\zeta_o^2}}.
     \label{FmaxVar}
\end{equation}
From Eqs. (\ref{DeltaVar}) and (\ref{FmaxVar}), we find \nn{the following: (i)  the value of $\delta \ell$ is independent of the width $ w_o $ of the strip. Further, in the vicinity of the buckling transition, $L\rightarrow L_o$, we observe that  (ii) $\delta \ell \propto \left| v\right|  \ell_o^2$,
and (iii)  $\left|\mathcal{F}_{max} \right| \propto  w_o \left|v\right| $.
The  dash-dotted lines in Fig. \ref{parameterVariation}  shows that the above  scaling laws  (i), (ii) and (iii) obtained from the three-spring model are consistent with our experimental findings}. It to be noted however that while this simple spring model is useful for determining scaling laws, it is incapable of capturing the buckling and mode coarsening phenomena observed in our experiments.

\section{Concluding remarks} \label{sec:conc}
In this work, we show that a free-standing soft thin object can be forced deep into a transiently compressed condition during dynamic loading. This causes mechanical instability, and the material exhibits negative compressibility for a brief period of time. This is determined by measuring in-plane forces and visualising the buckling process at the same time. During this process, the length and pressure change non-monotonically. Measuring these characteristics explains the growth and decay of the various observed modes.  The extent of this compression and the subsequent relaxation processes are determined by the narrow object's geometry and compression speed. We also show that the order of buckling and the forces generated in the system are controlled by compression speed. \\



\begin{acknowledgments}
We would like to thank Dr. H. Nanavati for useful discussions. We acknowledge support of the Department of Atomic Energy, Government of India, under Project No. 12-R\&D-TFR-5.10-0100.
\end{acknowledgments}


%

\end{document}